# Technically Plausible but Clinically Misleading? Expert Evaluation of Patient-Personalized Synthetic Prostate MRI

Gabriel Paulo Maglalang Israel[1], Sol Gedde[2], and Alvaro Fernandez-Quilez[1,2]

[1] Department of Electrical Engineering and Computer Science, University of Stavanger, Stavanger
[2] Stavanger Medical Imaging Laboratory (SMIL), Stavanger University Hospital, Stavanger
alvaro.f.quilez@uis.no

**Abstract.** Magnetic resonance imaging (MRI) is central to prostate cancer assessment, yet its acquisition is costly and time-consuming, making it a major bottleneck in the patient care pathway. Personalized MRI synthesis is a promising direction because it may allow the generation of clinically realistic images tailored to individual patients while reducing the dependence on scanner-acquired imaging. In this work, we investigate whether patient-personalized MRI synthesis produces images that experts perceive as clinically plausible and how such personalization influences expert interpretation. We synthesize patient-personalized prostate MRI using a 3D diffusion model conditioned on routine pre-imaging clinical variables age and PSA. Using a large publicly available dataset, we first verify that synthesized images are technically comparable to scanner-acquired MRIs using standard image similarity metrics. We then conduct a pilot, single-blinded, randomized expert study on a subset of cases sampled to reflect variation in patient clinical profiles. A radiologist compared personalized and unconditioned synthetic MRIs using paired assessments and Likert-scale ratings of clinical plausibility, confidence, and risk of misleading interpretation, supplemented by qualitative feedback. Although personalized synthetic MRIs appeared technically plausible, expert interpretation highlighted variability in confidence and potential risks of misleading cues when routine clinical data was embedded into image generation. These findings suggest that while personalized synthesis may be technically feasible, careful assessment is needed to understand how generated cues influence clinician interpretation before such systems can be safely integrated into clinical workflows.

## 1 Introduction

Prostate cancer (PCa) assessment begins with information that is routinely collected in clinical practice, such as patient age and prostate-specific antigen (PSA) levels [3]. These early signals are used to estimate cancer risk and determine whether more invasive or resource-intensive diagnostic procedures are warranted. One such procedure is MRI, which is commonly used to visualize prostate anatomy, identify suspicious regions, and guide prostate tissue extraction for diagnostic confirmation [4]. While MRI provides irreplaceable diagnostic information, its acquisition is costly, time-consuming, and not universally accessible. As a result, MRI often represents a bottleneck in the PCa patient care pathway, delaying downstream decisions and limiting the scalability of PCa assessment [16].

Recent advances in generative modeling have enabled highly realistic MRI synthesis, with diffusion-based models producing three-dimensional volumes with coherent anatomy [14]. Building on this foundation, prior work on medical image synthesis has sought to align generated images with patient-specific information by conditioning on signals such as segmentation maps, lesion annotations, or radiology reports [8, 7, 9]. While effective for controlling image appearance, these signals are often unavailable at early stages of care and may encode downstream diagnostic information that would not yet be available at that point in the patient pathway [20, 10, 17]. In contrast, routinely collected clinical variables such as patient age and PSA are available prior to MRI acquisition, consistently recorded, and directly inform early prostate cancer risk stratification, making them a natural basis for patient-centered personalization [6, 3]. Whilst conditioning on such variables may increase clinical feasibility, establishing whether such personalization is clinically meaningful depends on more than the ability to generate realistic images [6].

In clinical settings, the value of a generated image depends not only on its visual realism, but also on how it is interpreted by experts within a specific patient context [24, 2]. In such settings, MRI interpretation is shaped by expectations formed from available clinical signals, such as cancer risk indicators and referral context. This makes personalized synthesis a double-edged proposition: conditioning MRI generation on patient information may improve contextual plausibility, but it may also introduce cues that influence expert judgment in unintended or misleading ways [19, 16]. This suggests that assessing synthetic MRI involves not only visual realism, but also how generated images are perceived, trusted, and interpreted in relation to the patient information available at that stage of care [1, 11, 17].

*In our work, we distinguish between **perception**, referring to whether a synthetic image appears anatomically and clinically plausible for a given patient context, and **interpretation**, referring to how confidently the image can be reasoned about in context and whether it introduces risks of misleading clinical inference.*

We investigate how patient-personalized prostate MRI synthesis, conditioned on routine pre-imaging clinical variables such as age and PSA, influences expert perception and interpretation. We address the following research questions:

- **RQ1** How do experts perceive the plausibility of patient-personalized synthetic prostate MRI?
- **RQ2** How does conditioning prostate MRI synthesis on routine clinical data influence expert interpretation compared to unconditioned prostate MRI?
- **RQ3** What risks or failure modes do experts identify when interpreting patient-personalized synthetic prostate MRI?

## 2 Related Work

### 2.1 Personalization, Perception and Interpretation

Personalization is a common design strategy in clinical AI systems, where patient-specific variables such as demographics or laboratory measurements are used to tailor risk estimates or decision support [29, 12, 4]. In the broader clinical-AI literature, authors such as Vasey et al. have emphasized that early-stage evaluation must account for how AI systems are used and interpreted in practice, rather than relying on performance claims alone [12]. Most existing systems personalize symbolic or abstracted outputs (e.g., risk scores or alerts), which are interpreted through explicit reasoning about thresholds or probabilities [13, 22]. In contrast, as acknowledged in the work by Pickersgill et al., medical images are visually rich artifacts that rely on holistic and spatial assessment, making expert judgment sensitive to subtle perceptual cues [25, 23]. In that regard, routine pre-imaging clinical variables already inform how clinicians assess the plausibility and relevance of imaging findings [25]. However, it remains unclear whether including this information into the image generation process produces synthetic images that experts perceive as more patient-appropriate and how such perceptions shape subsequent interpretation. As Koetzier et al. note in their review of synthetic data for medical imaging, technical realism alone is not sufficient to establish clinical value [16]. Clarifying this distinction is therefore important for evaluating the practical clinical value of patient-personalized image synthesis [16].

### 2.2 Conditional Medical Image Synthesis and Evaluation

Recent work on generative medical imaging has shown that diffusion-based models can synthesize high-quality 3D images with coherent anatomy. In particular, Khader et al. demonstrated that denoising diffusion models can generate realistic 3D MRI and CT volumes based on technical and expert-based evaluations [14]. Building on this foundation, conditional generation has been used to control the anatomy or pathology of synthetic images through segmentation masks, annotations, reports, or prompts. For example, Konz et al. and Giardina et al. proposed segmentation-guided diffusion for anatomically controllable medical image generation, while Grabke et al. explored prompt-guided latent diffusion for 3D prostate MRI generation [8, 7, 18, 21]. While such conditioning strategies are effective for controlling image appearance, they often rely on signals that encode downstream diagnostic information unavailable prior to imaging [20, 28]. In contrast, routinely collected clinical variables such as age or PSA are available before imaging and already shape clinical expectations, but their use for conditioning image synthesis has received limited attention [16]. When used, the evaluation of the resulting synthetic medical images has primarily focused on technical fidelity or technical downstream task performance [7, 8]. At the same time, Koetzier et al. highlight that clinically relevant evaluation must also consider how synthetic images are interpreted and where they may be misleading in practice [16]. More broadly, work by Chambon et al. also underscores that clinical usefulness depends on alignment with domain context, not only generic visual performance [2]. However, few studies examine how conditioning generative models on routine clinical data affects expert evaluations of clinical plausibility and the perceived risk of misleading prostate MRI interpretation [20]. By focusing on expert perception and interpretation of personalized versus unconditioned synthetic prostate MRI, this work addresses this gap.

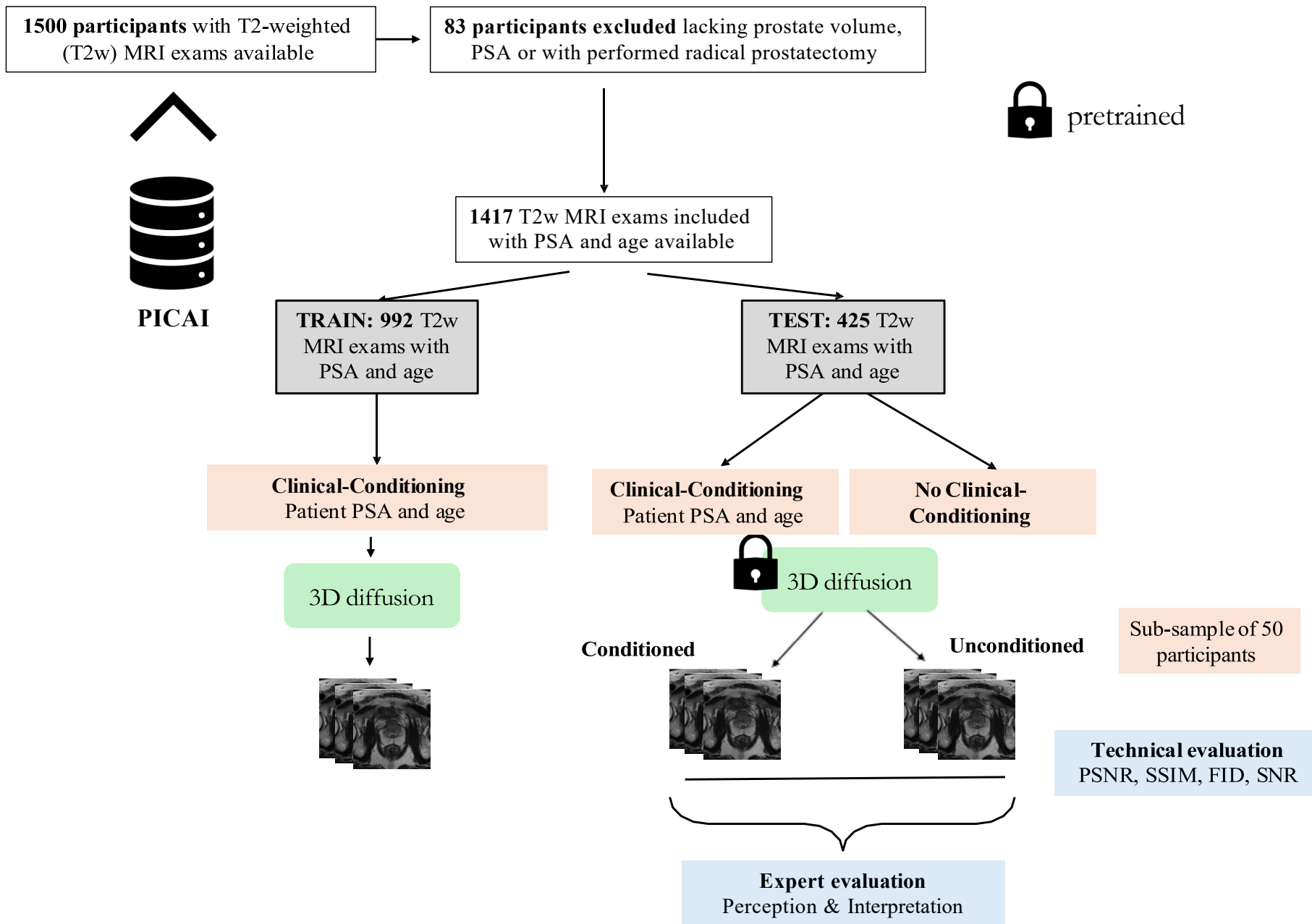

Fig. 1: Overview of the study methodology. Using the PI-CAI dataset, we trained unconditioned and patient-personalized 3D diffusion models for synthetic prostate MRI generation, where personalization was based on routine pre-imaging clinical variables (age and PSA). Synthesized images were first assessed through technical validation and then compared in a randomized expert evaluation to examine differences in perceived plausibility, interpretation confidence, and risk of misleading interpretation.

# 3 Methods

In the following, we describe the dataset, generative modeling approach, technical validation, and expert evaluation protocol used in this study. An overview of the approach is presented in Figure 1.

## 3.1 Dataset and Clinical Information

We use the publicly available PI-CAI dataset [27], a multi-center cohort of prostate MRI exams collected across four institutions and two scanner vendors between 2012 and 2021. The dataset includes T2-weighted MRI volumes and routinely collected clinical variables available during early prostate cancer assessment, including patient age and PSA.

We included imaging exams with complete clinical metadata and no prior prostate surgery, as prior intervention may alter prostate anatomy and thereby confound interpretation of synthesized images [27, 5]. This resulted in 1417 patient cases. T2-weighted MRI volumes were intensity-normalized and resampled to a common spatial resolution of $0.5 \times 0.5 \times 3.0$ mm to mitigate multi-scanner variability. The dataset was split at the patient level into 70% training (992 patients) and 30% test sets (425 patients) using stratified sampling to preserve distributions of clinical variables and diagnostic labels.

## 3.2 MRI Synthesis and Personalization

We implemented a 3D diffusion-based generative model to synthesize prostate T2-weighted MRI volumes. As a baseline, we trained an unconditional model using a latent-space denoising diffusion probabilistic model with a VQ-GAN encoder, following prior work on 3D medical image generation [14].

To enable personalization, we trained a conditional variant of the model to integrate routine clinical variables into the diffusion process. Patient age and PSA were encoded as textual prompts (e.g., "Patient is 65 years old and has a PSA value of 8.3 ng/mL") using a CLIP-based text encoder [26, 19]. This conditioning allowed the model to generate synthetic MRI volumes informed by patient-specific pre-imaging clinical context.

For each of the 425 patients in the test set, we generated two synthetic MRI volumes: (i) an **unconditioned** MRI generated without clinical conditioning, and (ii) a **personalized** MRI conditioned on the patient's age and PSA from the original test set. During synthesis, all other training and generation parameters were held constant so that differences between conditions could be attributed to the presence or absence of clinical conditioning.

## 3.3 Technical Validation of Synthesized Images

Prior to expert evaluation, we performed technical validation on the full test set. Personalized volumes were compared with the corresponding scanner-acquired

MRI from the same test case using PSNR, SSIM, and SNR, with results averaged across patients. For unconditioned volumes, no one-to-one patient correspondence exists by design. We therefore assessed similarity relative to the full test distribution by comparing each synthetic volume with all scanner-acquired test volumes and averaging the resulting scores. We additionally report Fréchet Inception Distance (FID) between the sets of synthetic and scanner-acquired volumes as a distributional realism measure [14]. This validation step was used to assess whether clinical conditioning was associated with systematic differences in image quality and whether synthesized images remained technically comparable across conditions at scale.

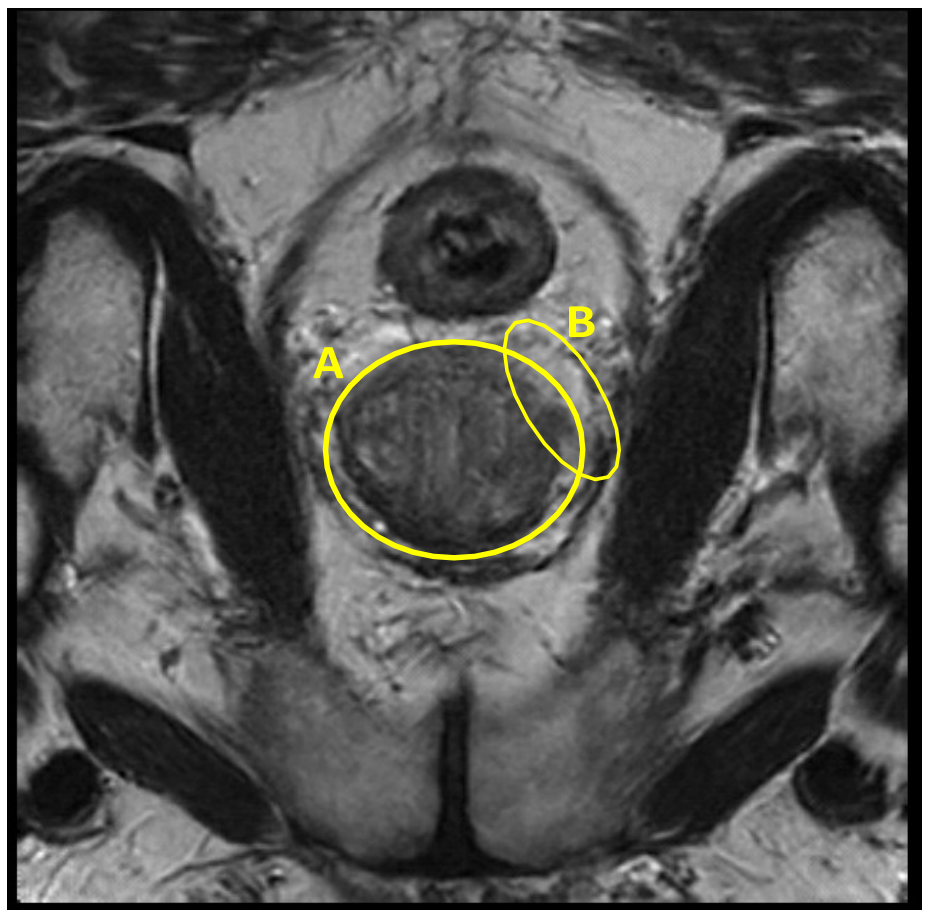

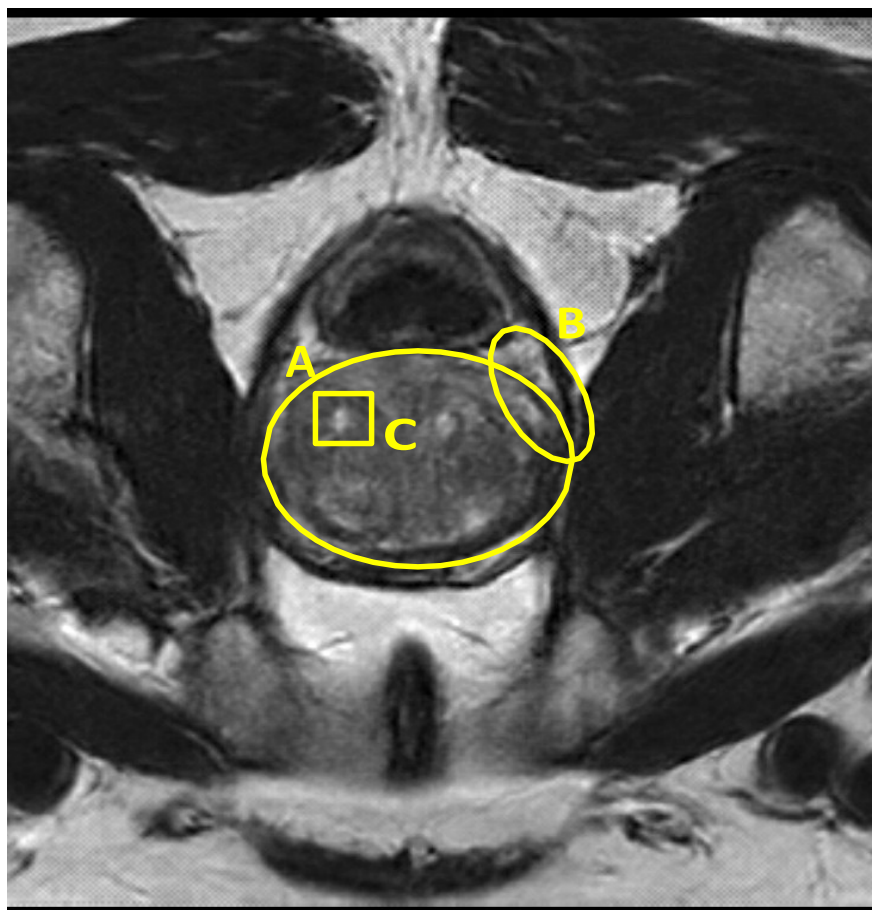

Fig. 2: Unconditioned (left) and patient-personalized (right) synthetic prostate MRI for a 58-year-old patient with a PSA value of 4.2 ng/mL. Highlighted regions correspond to cues noted by the radiologist, including (A) prostate capsule/boundary definition, (B) zonal anatomy, and (C) focal features that may encourage more specific interpretation.

### 3.4 Expert Evaluation

Once technically feasibility was evaluated, we conducted a pilot, single-blind, randomized, within-case expert evaluation to examine how clinical conditioning affects both expert perception and interpretation of synthetic prostate MRI. In this study, perception refers to judgments of anatomical and clinical plausibility, whereas interpretation refers to reasoning about how such images would be used in clinical practice and whether they may introduce misleading cues. The evaluation was designed to elicit these complementary dimensions.

**Participants, Stimuli and Procedure.** One radiologist with more than two years of experience in prostate MRI interpretation participated in the study. The expert evaluation was conducted on a subset of cases sampled from the technically validated test set of synthetic images. We sampled 50 patient cases from the test set to reflect variation in age and PSA. Cases were selected using stratified sampling across age and PSA (e.g., low/medium/high PSA and younger/middle/older age bins) to ensure coverage of clinically relevant ranges; the sample size was chosen to support repeated within-case comparisons while keeping the evaluation tractable. For each sampled case, the expert was presented with a paired comparison consisting of a personalized synthetic MRI and an unconditioned synthetic MRI. The participant was blinded to the MRI generation strategy. Presentation order was randomized, and images were blinded to the patient's diagnosis status. The corresponding patient age and PSA were provided with each case. To minimize expectancy effects, the expert was informed that the study examined differences between images, without being told during the evaluation which images were synthetic or how they had been generated.

The evaluation protocol was structured to separately capture (i) perceptual assessments of whether an image appears anatomically and clinically plausible for a given patient, and (ii) interpretive assessments concerning confidence in using the image for further clinical referral and the perceived risk that the image could mislead interpretation. For each conditioned-unconditioned pair, the expert completed:

1. A forced-choice selection of which MRI was more clinically plausible given the provided clinical context;
2. 7-point Likert-scale ratings for each MRI assessing:
   - anatomical consistency (perception),
   - clinical plausibility given age and PSA (perception),
   - confidence for clinical interpretation (interpretation),
   - risk of misleading interpretation (interpretation);
3. Free-text comments describing observed differences and potential concerns.

### 3.5 Analysis

Quantitative results are reported descriptively across cases, treating each case as a repeated observation from the single expert. We report descriptive statistics only (medians and IQRs for Likert items; proportions for forced-choice comparisons). Given the pilot design and single-expert sample, we do not perform inferential statistical testing.

Qualitative responses were analyzed using lightweight thematic analysis [15] to identify perceived benefits, risks, and failure modes of personalized image synthesis.

## 4 Results

### 4.1 Technical Validation of Synthesized Images

Prior to expert evaluation, we assessed the technical quality of synthesized images on the full test set (Table 1). Across paired similarity metrics, conditioning on age and PSA did not materially change PSNR, SSIM, or SNR relative to the unconditioned baseline. In contrast, conditioning substantially reduced FID, suggesting improved distribution-level realism while preserving overall image quality without introducing technical artifacts nor MRI degradation.

Table 1: Image quality metrics for unconditioned (baseline) and patient-personalized synthetic prostate MRI (age/PSA-conditioned). Values are averaged across diagnostic strata and reported as mean ± std. Arrows indicate direction of improvement.

| Metric | Unconditional (baseline) | Personalized |
|---|---|---|
| PSNR (↑) | 12.14 ± 0.07 | 12.06 ± 0.10 |
| SNR (↑) | 1.37 ± 0.01 | 1.34 ± 0.01 |
| SSIM (↑) | 0.11 ± 0.01 | 0.11 ± 0.01 |
| **FID (↓)** | 1.05 ± 0.04 | 0.68 ± 0.04 |

### 4.2 Expert Perception: Plausibility of Patient-Personalized Images

Across the 50 evaluated cases, personalized images were more frequently selected as the more clinically plausible representation for the given patient context (Table 2). Forced-choice comparisons showed that the conditioned image was preferred in 76% of cases, suggesting that personalization often made the generated MRI appear more appropriate for the provided age and PSA profile.

Likert-scale ratings showed the same overall pattern. Personalized images received higher median ratings for both anatomical consistency and clinical plausibility than unconditioned images. The largest differences were observed on plausibility-related dimensions, indicating that conditioning on routine clinical variables primarily affected how well the image appeared to fit the stated patient context, rather than simply making the image look generically more realistic. In practice, this was reflected in judgments about clearer zonal anatomy, more coherent gland boundaries, and a stronger sense that the image "matched" the accompanying patient information.

### 4.3 Expert Interpretation: Confidence, Risk, and Failure Modes

Median ratings for confidence in clinical interpretation were modestly higher for personalized images than for unconditioned images (Table 2). This suggests that conditioning sometimes made the images easier to reason about in relation to the provided patient context. However, the increase in confidence was smaller than the increase in perceived plausibility, indicating that images judged as more plausible were not automatically regarded as straightforward to interpret clinically.

Personalized images were associated with a higher perceived risk of misleading interpretation in a substantial minority of cases. Specifically, the perceived risk rating increased for conditioned images in 36% of within-case comparisons. Figure 2 illustrates this trade-off in a representative case. In this example, clearer anatomy and zonal structure appeared to increase plausibility, while localized focal cues invited a more specific interpretation and raised concern about whether the image might be read too confidently relative to the limited conditioning information used to generate it.

Table 2: Expert evaluation summary for clinically conditioned versus unconditioned synthetic prostate MRI ($n = 50$ cases). Likert ratings are median [IQR] on a 7-point scale (1 = strongly disagree, 7 = strongly agree).

| **Measure** | **Conditioned** | **Unconditioned** |
|---|---|---|
| Preferred as more plausible (%) | 76 | 24 |
| Anatomical consistency | 5 [4–6] | 3 [3–4] |
| Clinical plausibility | 5 [4–6] | 3 [2–4] |
| Interpretation confidence | 4 [3–5] | 3 [3–4] |
| Perceived risk of misleading interpretation | 4 [3–5] | 3 [2–4] |
| Risk increased (cases %)† | 36 | – |

†Risk increased denotes the proportion of cases where the perceived risk rating for the conditioned image was higher than for the unconditioned image (within-case). For all items, higher values indicate stronger endorsement of the statement; for the risk of misleading interpretation item, higher values correspond to greater perceived risk.

### 4.4 Expert Feedback: Qualitative Themes

Based on the thematic analysis[3], three interrelated and recurring themes were observed across cases: (*i*) improved perceptual plausibility, (*ii*) expectation anchoring to patient context, and (*iii*) risk of over-specific or misleading cues. These themes were not mutually exclusive. In several cases, the same features that increased plausibility also appeared to make the image more persuasive and therefore potentially more misleading during interpretation.

***Improved perceptual plausibility.*** The first and most consistent theme concerned the visual and anatomical coherence of personalized images. Comments in this category often referred to clearer zonal anatomy, more coherent capsule or gland boundaries, and a stronger overall sense that the image resembled a clinically plausible prostate MRI. The radiologist's feedback suggested that conditioned images more often appeared internally consistent and easier to accept as a reasonable patient-specific representation. This theme aligns closely with the higher forced-choice preference for conditioned images and the increased ratings for anatomical consistency and clinical plausibility.

> *"The zonal anatomy appears more coherent, which makes the image feel more clinically plausible overall."*

[3] Blue highlights refer to perceptual plausibility (visual and anatomical coherence); orange highlights indicate expectation anchoring to patient context; red highlights denote perceived interpretive risk or potential for misleading cues.

In several cases, the expert's comments suggested that this increased plausibility was not limited to one isolated region, but reflected a more global impression of anatomical organization. Put differently, personalization often appeared to improve the "readability" of the synthetic image at first glance, making it feel more consistent with expectations of how prostate MRI should look.

***Expectation anchoring to patient context.*** A second recurring theme concerned the way patient age and PSA shaped interpretation before or during image assessment. The expert did not assess the image in isolation, but instead appeared to reason about it in relation to what would be expected for the provided patient profile. In this sense, the conditioning information functioned not only as model input, but also as interpretive context. When the personalized image aligned with those expectations, it was more likely to be judged as plausible. When alignment was weaker, the unconditioned image could sometimes appear less obviously wrong, even if it was less tailored.

> *"Given the patient's age and PSA, I would expect certain signal characteristics, and one image aligned more closely with those expectations."*

This theme helps explain why personalization altered expert assessment even when the conditioning variables themselves were relatively limited. Age and PSA do not specify lesion location or pathology directly, but they do shape the background expectations against which image features are judged. The qualitative feedback supported the idea that personalization affects not only MRI appearance, but also the interpretive frame brought to the MRI by the clinician.

***Risk of over-specific or misleading cues.*** The third theme captured concerns that personalized synthesis could introduce features that appeared more specific than warranted by the available conditioning information. In these cases, the expert noted focal structures, signal changes, or localized cues that invited interpretation as clinically meaningful findings, even though such specificity could not be justified by age and PSA alone. This theme was especially important because it often co-occurred with higher plausibility: images that looked more coherent or better matched to the provided context could also feel more persuasive, thereby increasing the risk of over-interpretation.

> *"If the image appears too strongly shaped by age or PSA, there is a risk of over-interpreting findings that are not actually supported."*

The radiologist's comments suggested two related failure modes within this theme. First, personalization could encourage ***anchoring*** , where an apparently plausible feature became overly influential simply because it seemed to fit the provided patient context. Second, personalization could imply an unwarranted degree of ***specificity*** , making synthetic cues seem more diagnostically meaningful than they were.

## 5 Discussion

This work examined how conditioning synthetic prostate MRI on routinely collected clinical variables shapes expert perception and interpretation. In particular, we investigated how patient-personalized synthesis compares to unconditioned generation and what implications this has for clinical reasoning. Prior work on conditional medical image synthesis has primarily emphasized technical fidelity or technical downstream task performance, often using conditioning signals unavailable prior to prostate MRI imaging [18, 8]. Our findings extend broader concerns that technically plausible and clinically realistic synthetic outputs may still shape expert interpretation in potentially misleading ways [16].

Personalized images were consistently perceived as more clinically plausible than unconditioned images (**RQ1**). In forced-choice comparisons, the conditioned image was preferred in most cases, and median ratings for both anatomical consistency and clinical plausibility were higher. Qualitative feedback suggests that these judgments were often tied to clearer zonal anatomy, better boundary definition, and a stronger overall sense of coherence. These findings indicate that conditioning on age and PSA can make synthetic prostate MRI appear more appropriate for a given patient context. In this respect, our results extend prior work by Khader et al. and Grabke et al.: whereas those studies primarily show that diffusion-based synthesis can produce realistic or task-useful images conditioned on non-routinely available clinical data, our findings suggest that personalization using routine clinical variables can also increase perceived patient-specific plausibility during expert review [14, 8].

Beyond absolute plausibility, **RQ2** concerns how conditioning changes expert judgment relative to unconditioned synthesis. Our results suggest that personalization does not merely affect the clinical realism of MRI, but actively shifts assessment when the MRI is interpreted in context. Compared with unconditioned outputs, personalized MRI were more often judged as plausible and received somewhat higher confidence ratings. The qualitative data help explain this effect: age and PSA shaped the interpretive frame through which the expert evaluated the image. In this sense, personalization functioned not only as a generative mechanism, but also as a contextual cue that influenced how image features were read. This distinguishes our contribution from prior prostate MRI synthesis studies such as Grabke et al., where conditioning is mainly treated as a way to improve controllability or downstream model performance; here, conditioning also altered how the generated image was understood by a clinician [8].

At the same time, personalization introduced new interpretive dynamics (**RQ3**). In 36% of cases, conditioning increased the perceived risk of misleading interpretation. Thematic analysis suggests that this occurred when contextual fit and focal MRI cues combined to create an impression of specificity that exceeded what the input variables could reasonably support. In this sense, the same properties that enhanced plausibility also appeared to contribute to interpretive risk. Personalized prostate MRI could therefore become more convincing without necessarily becoming safer to interpret. This observation resonates with the broader argument by Koetzier et al. that synthetic medical images should not be evaluated solely in terms of fidelity or utility, but also in terms of how they may be used, trusted, and potentially misused in clinical workflows [16].

These findings emphasize that conditioning on routine clinical variables can make generated MRI feel more contextually appropriate and easier to accept as plausible clinical artifacts. However, the same contextual fit may also make synthetic MRI more persuasive, increasing the chance that experts attribute undue meaning to generated cues. In this sense, personalization appeared to operate not only as a technical conditioning mechanism, but also as an interpretive framing mechanism. This has implications for how patient-personalized generative systems should be evaluated and potentially clinically used. Technical fidelity and expert reader preference alone are not sufficient to establish clinical value. As suggested both by our findings and by broader reviews of synthetic medical imaging, evaluation should also examine how personalization changes what clinicians attend to, how confidently they reason from generated images, and under what circumstances plausibility and perceived risk diverge [16]. Such questions are especially important when conditioning variables are available early in the care pathway and may shape expectations before diagnostic imaging is acquired.

This study is exploratory and limited to a single expert, and the reported results should therefore be interpreted as formative rather than generalizable. The use of one radiologist allowed for close within-case comparison and detailed qualitative feedback, but it does not capture variability across levels of expertise, interpretive style, or institutional practice. Similarly, the study focused on age and PSA as examples of routinely available pre-imaging variables; other forms of patient information may produce different effects on image plausibility and interpretive risk. Future work should therefore extend this analysis to multiple clinicians, additional clinical variables, and different points along the patient care pathway. More broadly, our findings suggest that the next step for patient-personalized synthesis is not only to improve realism or controllability, but also to develop evaluation frameworks that capture how synthetic images shape clinical reasoning in context.

**Acknowledgments.** The authors thank the organizers of the PI-CAI challenge for making the dataset publicly available and for their efforts in curating a large, multi-center prostate MRI cohort.

**Disclosure of Interests.** The authors have no competing interests.